\documentclass[10pt,twocolumn]{article} 

\usepackage{times}
\usepackage{graphicx}
\usepackage{amssymb}
\usepackage{url,hyperref}
\usepackage{natbib}

\usepackage{textcomp}
\usepackage{rotating}
\usepackage{pdflscape}

\begin{document}

\title{GeneVis\\
\large An interactive visualization tool for combining cross-discipline datasets within genetics}
\author{Casper van Leeuwen\\
\\
Visualization group, SURFsara, Amsterdam, the Netherlands.\\
\\
\href{casper.vanleeuwen@surfsara.nl}{casper.vanleeuwen@surfsara.nl}\\}

\maketitle

\abstract{
\textbf{Summary:} GeneVis is a web-based tool to visualize complementary data sets of different disciplines within the field of genetics. It overlays gene-cluster information, gene-interaction data and gene-disease association data by means of web-based interactive graph visualizations. This allows an intuitive and quick assessment of possible relations between the different datasets. By starting from a high-level graph abstraction based on gene clusters, which can be selected for detailed inspection at the gene-interaction level in a separate window, GeneVis circumvents the common visual clutter problem when using gene datasets with a high number of gene entries. \\
\textbf{Availability:} The source of GeneVis is available at \href{https://gitlab.com/caspervanleeuwen/genevis/}{https://gitlab.com/caspervanleeuwen/genevis/}\\
\textbf{Contact:} \href{casper.vanleeuwen@surfsara.nl}{casper.vanleeuwen@surfsara.nl}\\
}

\section{Introduction}

The continuous technological advancement in the fields of genetics, genomics and bioinformatics has led to large data sets on gene clusters, gene-gene interactions and the association of genes with diseases. One of the challenges is to use this wealth of information by combining the data from different sources and experiments to get a better (intuitive) understanding of the biological processes underpinning disease. GeneVis is a web-based visualization tool to explore high-level gene clusters (e.g. Kyoto Encyclopedia of Genes and Genomes\footnote{http://www.kegg.jp/}, KEGG, pathway functions), gene-to-gene interactions (e.g. Human Integrated Protein-Protein Interaction Reference\footnote{http://cbdm-01.zdv.uni-mainz.de/~mschaefer/hippie/}, HIPPIE) and gene-disease associations (e.g. Genome-Wide Association Studies\footnote{http://www.gwascentral.org/}, GWAS). Data from these sources are generated in a domain-specific way (i.e in their own format using different methods), which makes it difficult to directly relate data from the different sources to each other. GeneVis was designed to overcome this problem by generating insightful visualizations to create a higher-level understanding of multi-domain research results.

\begin{sidewaysfigure*}[p]
    \centering
    \includegraphics[keepaspectratio=true,width=\linewidth]{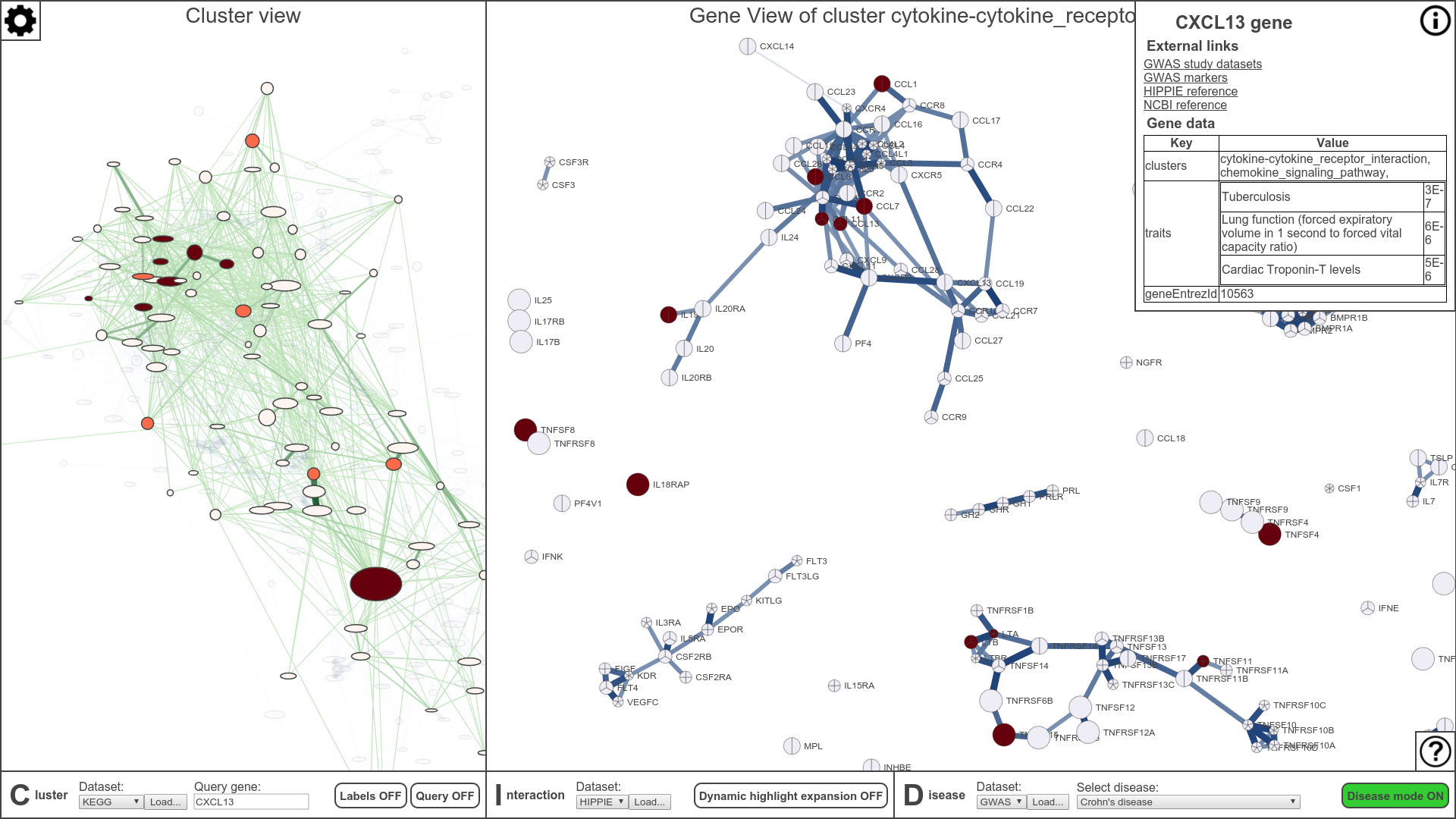}
\caption{GeneVis. This image is the full interface of the GeneVis. The left panel is the Cluster View and the right panel is the Gene View. The bottom of the interface is the interaction bar, which contains all the data set loading and graph interaction buttons.}\label{fig:02}
\end{sidewaysfigure*}

GeneVis uses force-directed graphs as a basis to visualize relations between gene clusters, between clusters and individual genes and between individual genes (figure~\ref{fig:02}). Force-directed graphs are a general way to visualize relational data~\citep{eades}\citep{tollis}\citep{thomas}, although their effectiveness can succumb to visual clutter whenever the number of data items gets large. To reduce this visual clutter the GeneVis visualization tool introduces a higher-level view in the form of a cluster graph, called the Cluster view, where sets of genes are used as graph nodes instead of individual genes (left panel, figure~\ref{fig:02}). This reduces the number of nodes in the graph and reduces the number of visual items to inspect and comprehend. This higher-level cluster graph gives the user a concise but insightful view of the whole gene dataset, using node deformations and color maps to display the average gene association to the cluster and cluster size and the average association with a disease, respectively (left panel, figure~\ref{fig:02}). GeneVis also gives the user the ability to "zoom" into a cluster of interest and reveal the underlying gene subset and gene interactions. This level is called the Gene View (right panel, figure~\ref{fig:02}). Both in Cluster and Gene View gene sets can be explored by zooming, selecting and highlighting genes or clusters. Finally, a visual-scheme (or color map) over both views called the "Disease Mode" allows quick visual inspection of the association of genes or gene clusters with a specific disease or trait, using data from gene-disease association datasets (GWAS data by default).

\begin{figure}
    \centering
    \includegraphics[width=\columnwidth]{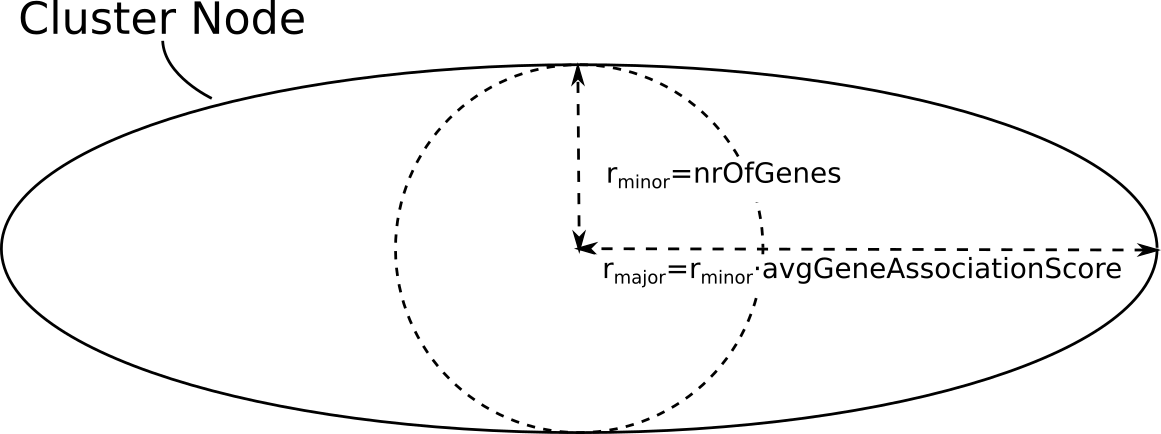}
\caption{Cluster node representation. This shows that the minor radius (radius inner circle) is defined by the number of genes and the major radius is defined by the minor radius multiplied by the average gene association score.}\label{fig:01}
\end{figure}

\section{Method}
\label{secMethod}

\subsection{Datasets}
\label{secData}
GeneVis allows a simultaneous visualization of three complementary types of datasets: a gene cluster dataset, a gene-gene interaction dataset and a gene-disease association dataset. All datasets must be provided in either tab or comma-separated tabular file formats. For example file formats see Appendix .

\subsubsection{Gene cluster dataset}
To visualize gene clusters with GeneVis, the gene cluster dataset should contain a list of gene names with the associated gene cluster. The cluster data can either be based on a hard (0 or 1) or soft probability clustering (between 0 and 1), which both will be handled appropriately by GeneVis. Each cluster must contain a subset of genes, where each gene has an association value ${a_c}$, where $c$ is the cluster to which the gene is associated. This association value ${a_c}$ can either be a given $p$-value from a dataset (for a soft-clustered dataset) or an automatically computed fraction ${a_c=\frac{1}{N^c_g}}$(for a hard-clustered dataset), determined by the number of clusters ${N^c_g}$ a specific gene is associated with. Each row within the dataset, except the first header row, represents a gene. The columns represent the gene Entrez ID (geneEntrezId), the gene name (geneName) and all the gene clusters. All the columns must be named accordingly in a header row. The association values between genes and clusters are stored in the cells of the tabular structure (see table ~\ref{geneclusterformat}).

\begin{table}
\centering
\resizebox{\columnwidth}{!}{%
\begin{tabular}{|l|l|c|c|c|c|}
\hline
\textbf{geneEntrezId} & \textbf{geneName} & \textbf{GLYCOLYSIS ...} & \textbf{CITRATE\_CYCLE ...} & \textbf{PENTOSE ...} & \textbf{...} \\
\hline
873            & CBR1 & 0.2                                 & 0.4                              & 0.9                     & ...          \\
\hline
2026            & ENO2 & 0.6                                 & 0.6                              & 0.2                      & ...         \\
\hline
2665           & GDI2 & 0.1                                 & 0.2                              & 0.1                            & ...   \\
\hline
...           & ... & ...                                 & ...                              & ... & ... \\
\hline                          
\end{tabular}%
}
\caption{Gene cluster dataset format example}
\label{geneclusterformat}
\end{table} 

\subsubsection{Gene-gene interaction dataset}
GeneVis can visualize any form of gene-gene interaction that can be scored as an interaction strength between 0 and 1. To visualize gene-gene interactions, GeneVis requires a tabular dataset of genes containing 3 columns, the source gene Entrez ID (SourceGeneId), the target gene Entrez ID (TargetGeneId) and the interaction score (score)(see table ~\ref{genegeneformat}). All the columns must be named accordingly in a header row.  Each row entry contains the interaction strength (score, varying between 0-1) between 2 individual genes (source and target), which will affect the representation in Gene View, see section~\ref{secGeneView}. Different types of gene-gene interactions can be visualized in this way, for instance protein-protein interactions discovered with co-immunoprecipitation assays or yeast-2-hybrid~\citep{stelzl}\citep{likw}. 

\begin{table}
\centering
\begin{tabular}{|c|c|c|}
\hline
\textbf{SourceGeneId} & \textbf{TargetGeneId} & \textbf{score} \\
\hline
216          & 216          & 0.75  \\
\hline
3679         & 1134         & 0.73  \\
\hline
55607        & 71           & 0.65  \\
\hline
...          & ...          & ...  \\
\hline
\end{tabular}%
\caption{Gene-gene interaction dataset format example}
\label{genegeneformat}
\end{table}

\subsubsection{Gene-disease association dataset}
The gene-disease association dataset is a tabular dataset containing diseases and its associated genes and scores. Each row in this dataset is a study about a certain disease and it's associated gene(s) and that studies $p$-value (see table ~\ref{genediseaseformat}). This dataset must contain at least the following columns: gene names (Genes), disease/trait (Disease/Trait) and the p-value of the association ($p$-value).  This dataset will be visually represented in the Disease Mode, see section~\ref{secDiseaseMode}.

\begin{table}
\centering
\resizebox{\columnwidth}{!}{%
\begin{tabular}{|l|l|c|c|}
\hline
\textbf{Disease/Trait}       & \textbf{Genes}          & \textbf{p-Value}     & \textbf{...} \\
\hline
depressive disorder & CBX4           & 0.0000002   & ... \\
\hline
depressive disorder & PDZD2          & 0.0000003   & ... \\
\hline
depressive disorder & CTC-497E21.5   & 0.0000007   & ... \\
\hline
...                 & ...            & ...         & ... \\
\hline
\end{tabular}%
}
\caption{Gene-disease association dataset format example}
\label{genediseaseformat}
\end{table} 

\subsection{Cluster View}
\label{secClusterView}

GeneVis offers two levels of visualization. In Cluster View (left panel, figure~\ref{fig:02}) all gene clusters present in the gene cluster dataset and their overlap are shown. The layout of the graph of clusters is determined by means of a force-directed graph layout algorithm~\citep{tollis}. Each gene cluster is represented with an ellipse-shaped node, where the ellipse size defined by the inner circle radius represents the number of genes present in the cluster and the length of the major axis of the ellipse is inverse proportionally scaled to represent the average of gene association scores within the cluster multiplied by the minor radius(figure~\ref{fig:01}). So spherical clusters have a higher average gene association score than more elliptical clusters and proportionally smaller nodes have less associated genes than proportionally bigger nodes.  A random color is assigned to each node and while the Disease Mode is active they are colored according to their disease association, see section~\ref{secDiseaseMode} and figure~\ref{fig:02} and figure~\ref{fig:03}.

An edge between two cluster nodes describes the gene overlap between the clusters. The size and the color intensity of the edge both scale with the amount of overlap between the clusters, meaning the wider and brighter the edge, the more overlap there is between the connected clusters, in terms of numbers of genes. 

Clusters can be inspected by highlighting the cluster, which will reveal the underlying genes associated with the cluster in GeneView.

\subsection{Gene View}
\label{secGeneView}

The Gene View (right panel, figure~\ref{fig:02}) visualizes the individual gene interactions within a selected cluster. Each gene is represented as a node and each edge is an interaction between the genes. The gene nodes are represented as pie charts, where the size represents that gene’s association with the selected cluster and the pie pieces represent all the clusters in which that gene is present, proportional by cluster membership. The edge width and color represents the interaction score between genes. 
The Gene View has an extensive interaction model for the user to explore the gene datasets. Individual genes can be highlighted and queried by hovering and clicking, respectively. There are three methods to highlight the gene interactions:  

1) "Level of connectedness highlighting" which highlights all the gene interactions connected to the highlighted gene through different levels of connectivity, e.g. a level of connectedness of 2 reveals all the connected genes to the highlighted gene as level 1 and all the genes connected to the genes in level 1 as level 2.  

2) "Link threshold highlighting" progresses through all the connected gene interactions and highlights those interactions whose interaction score is above the given threshold.  

3) "Top-n link highlighting" highlights the top n highest interaction links and its genes starting from the highlighted gene.

\subsection{Disease Mode}
\label{secDiseaseMode}

The Disease Mode will change the color and opacity representation of the nodes and edges of both the Cluster View and the Gene View based on a selected disease from the gene-disease association dataset.

The Cluster View cluster nodes and edges that do not contain any genes associated to the selected disease will have a lower opacity in addition to their color and size. This reveals a disease sub-cluster graph network, which gives a sparse overview on the disease and cluster relations. Each cluster node will be colored according to that clusters enrichment score, where highly significant scores (${p<0.05}$) appear red, weakly significant scores (${0.05<p<0.1}$) appear orange and not significant scores (${p>0.1}$) appear white. This enrichment score is a modified one-tailed Fisher Exact $p$-value called the EASE score.

The Gene View gene nodes are color-coded according to the $p$-values associated to the study behind the selected disease obtained from the gene-disease association dataset.

\section{Application}
\label{secApplication}
\begin{figure*}[p]
    \centering
    \includegraphics[keepaspectratio=true,height=\textheight]{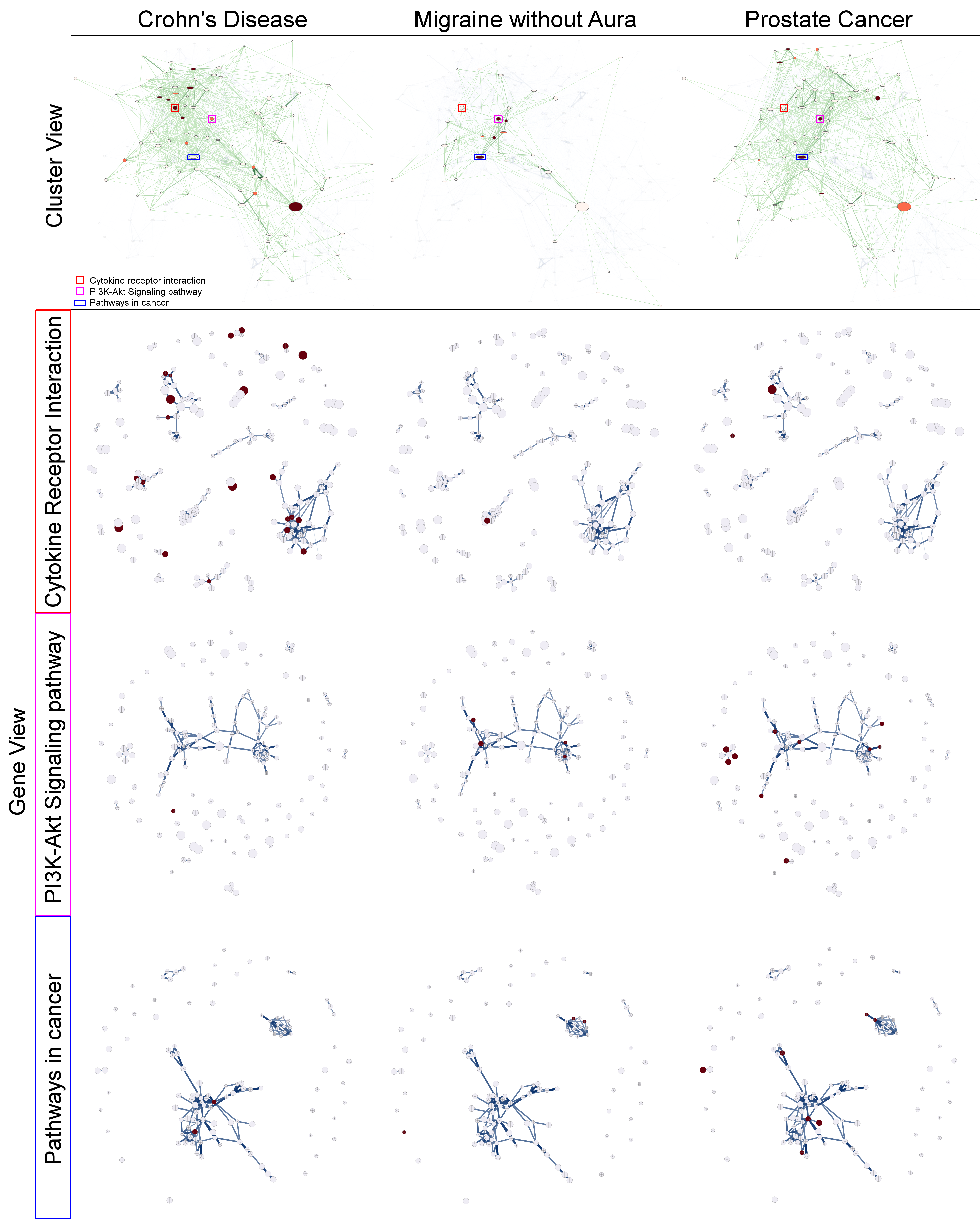}
\caption{The Cluster View (first row) and the Gene View (last 3 rows). The first row shows a comparison of 3 Cluster Views representations for 3 different diseases: Crohn's Disease, Migraine without Aura and Prostate Cancer. In the Cluster View row the three different pathways are annotated to show the comparison between these pathways relative to the selected disease. For example, the Cytokine receptor interaction pathway is an active pathway in inflammatory diseases and as Crohn's disease is an inflammatory bowel disease it has a significant enrichment score and is color coded accordingly in red. The 3 Gene View rows show 9 comparative GeneView Disease Mode representations between the 3 different diseases and 3 different pathways. Just like the Cluster View a clear difference can be seen between the diseases in the different pathways.}\label{fig:03}
\end{figure*}

In figure~\ref{fig:02} and~\ref{fig:03} an example of the tool is given for the gene cluster dataset KEGG, gene-gene association dataset HIPPIE and gene-disease association dataset GWAS. Figure~\ref{fig:02} shows the UI of GeneVis with on the left side panel the clusters and the right side panel the genes associated to the Cytokine receptor cluster. Both are colored coded by their association to the Crohn's disease(where red is ${p<0.05}$, orange is ${0.05<p<0.1}$ and white is ${p>0.1}$). Figure~\ref{fig:03} shows on the first row an in depth comparison of the cluster view color mapped with 3 different diseases, Crohn's disease, Migraine without Aura and Prostate cancer. The following 3 rows show an in depth comparison of the gene view of 3 different clusters (Cytokine receptor interaction, pathways in cancer and CPI3k-Akt signaling pathway) color mapped by the same diseases.  In this example Genevis shows which of the clusters are associated to the currently selected disease. These gene-cluster and disease association can than be further explored on a gene-level in the Gene View. In turn the Gene View shows which genes, which have not been associated to the disease yet, are connected to the disease associated genes. This way the user can exploratively find potential new studies when it comes for example gene-disease associations.

\section{Implementation}
\label{secImplementation}

GeneVis is a web-based visualization tool written in HTML, CSS, JavaScript, jQuery\footnote{http://jquery.com/} and the framework D3.js\footnote{http://d3js.org/}. Roughly \texttildelow80\% of the application uses the framework D3.js, which is a versatile data-driven HTML DOM element manipulation framework. D3.js made it possible to provide user's with full interactive control over their gene cluster data with interactive frame rates.

\begin{thebibliography}{}

\bibitem[Eades {\it et~al}., 1984]{eades}
Eades, P. (1984) A heuristic for graph drawing, {\it Congressus Numerantium}, {\bf 42}, 142-160.

\bibitem[Tollis {\it et~al}., 1999]{tollis}
Tollis, I.G., Battista, G.D., Eades, P., and Tamassia, R. (1999) Graph drawing - Algorithms for the visualization of graphs, Prentice Hall, Upper Saddle River, NJ.

\bibitem[Fruchterman and Reingold, 1991]{thomas}
Thomas M. J. Fruchterman and Edward M. Reingold (1991) Graph Drawing by Force-directed Placement, {\it Software-Practice and Experience}, {\bf 21}, 1129-1164.

\bibitem[Hosack {\it et~al}., 2003]{hosack}
Hosack, D. A. , Dennis, Jr. G., Sherman, B. T., Clifford Lane, H., Lempicki, R. A. (2003) Identifying biological themes within lists of genes with EASE Genome Biol, {\it Genome Biol.}, {\bf 4}:R70.

\bibitem[Stelzl {\it et~al}., 2005]{stelzl}
Stelzl, U., Worm, U., Lalowski, M., Haenig, C., Brembeck, F.H., Goehler, H., Stroedicke, M., Zenkner, M., Schoenherr, A., Koeppen, S., Timm, J., Mintzlaff, S., Abraham, C., Bock, N., Kietzmann, S., Goedde, A., Toksöz, E., Droege, A., Krobitsch, S., Korn, B., Birchmeier, W., Lehrach, H., Wanker, E.E. (2005) A human protein-protein interaction network: a resource for annotating the proteome, {\it Cell}, {\bf 122(6)}, 957-68.

\bibitem[Li {\it et~al}., 2010]{likw}
Li, K.W., Klemmer, P., Smit, A.B. (2010) Interaction proteomics of synapse protein complexes, {\it Anal Bioanal Chem.}, {\bf 397(8)}, 3195-202.


\end{thebibliography}
\end{document}